\UseRawInputEncoding
\documentclass[preprint,12pt]{elsarticle}%
\usepackage{amssymb}
\usepackage{amsfonts}
\usepackage{amsmath}
\usepackage{graphicx}%
\providecommand{\U}[1]{\protect\rule{.1in}{.1in}}
\graphicspath{{E:/dottorato/Coulomb_damper/Testo/Paper/Immagini/}}
\journal{journal}

\begin{document}
%
\begin{frontmatter}%


%

\title
{A comment on "Discussion on the use of the strain energy release rate for fatigue delamination characterization"}%

%

\author{M. Ciavarella (1), A. Papangelo (1), G.Cricrì (2)}%
%

\address
{(1) DMMM department. Politecnico di BARI. Viale Gentile 182, 70126 Bari. Italy. Mciava@poliba.it
(2) Università di Napoli Federico II, DII dept., P.Tecchio, 80. 80125 Napoli (It)}%
%

\begin{abstract}%

In a recent very interesting and illuminating proposal, Yao et al. (2014) have
discussed the use of the strain energy release rate (SERR) as a parameter to
characterize fatigue delamination growth in composite materials. They consider
fatigue delamination data strongly affected by R-curve behaviour due to fibres
bridging and argue that a better approach is to correlate the crack advance
with the total work per cycle measured in the testing machine. This seems to
work better than estimating the compliance as a linear fit of experimental
curves from Modified Compliance Calibration ASTM standards equations for the
SERR in the classical Linear Elastic Fracture Mechanics framework. We show
however that if we assume indeed linear behaviour (i.e. LEFM), the approach
they introduce is \textit{perfectly equivalent} to the SERR one, i.e. Paris
type of laws. As well known form Barenblatt and Botvina, fatigue crack growth
is a weak form of scaling, and it gives Paris classical dependence only when
the crack is much longer than any other characteristic sizes. Paris' law is
not a fundamental law of physics, is not an energy balance equation like
Griffith, and strong size effects due to cohesive zones have been found
already in concrete by Bazant. The proposal is very simple, and interesting as
it would seem to suggest that a proper scaling with a cohesive model at crack
tip could be predicted, although this doesn't seem to have been attempted in
the Literature. The main drawback of the present proposal is that it is not
predictive, but purely observational, as it requires the actual measurement of
work input during the fatigue process.%

\end{abstract}%
%

%

\end{frontmatter}%



\section{\bigskip Introduction}

The paper by Yao et al. (2014) discuss that the use of the strain energy
release rate SERR ($G$) to characterize fatigue delamination growth, either in
the form of its maximum $G_{\max}$, or in the range $\Delta G$ is problematic
in delamination, because of crack bridging, which leads to a crack R-curve in
static tests, and therefore quite intuitively a shift of Paris curves also
during fatigue loading. They find moreover that fatigue precrack or static
precrack lead to different fibre bridging, so they argue that bridging is
fundamentally different in static or fatigue conditions. But why they say the
SERR is not an appropriate parameter?

Their proposal is to correlate the crack advance per cycle $da/dN$ with
$dU/dN$, where $U$ is applied work per cycle. If we limit our discussion to
the case of loading ratio $R=0$ (althogh their paper contains experiments at
$R=0.5$), we can define as
\begin{equation}
U=\frac{1}{2}P_{\max,N}\delta_{\max,N}%
\end{equation}
where $P_{\max,N}$ is the maximum force at the cycle number $N$; $\delta
_{\max,N}$ is the maximum displacement at the cycle number $N$. Therefore, $U$
is computed "with the loads and displacements measured during the fatigue
tests" -- unfortunately, there are no plots of the load vs displacement
relationship to check whether they are really linear so as to use this simple
equation, rather than a full integration of the load-displacement curve. We
use force per unit thickiness, so also energies are per unit thickness and we
simplify the notation.

As they write for their DCB specimen, their eqt.2 shows that the SERR
\begin{equation}
G_{\max}\propto P_{\max}^{2}C^{2/3} \label{DCB}%
\end{equation}
where $C$ is the compliance. Also, as they say and as the ASTM D5528-01
standard suggests, the compliance is proportional to%
\begin{equation}
C\propto a^{3}%
\end{equation}
where $a$ is crack size. $P_{\max}$ varies during their test, which control
the $d_{\max}$.

It seems therefore that they are essentially using an estimate of SERR which
comes from a LEFM model, although it would seem that fibre bridging would make
the problem non linear, and hence in using the ASTM equation for the
compliance, there is an intrinsic simplification of the SERR. Unfortunately,
they don't show also the compliance measurement, to judge how this is linear,
and whether the use of ASTM Modified Compliance Calibration really is a
linearized curve fit about some intermediate condition.

\bigskip A general calculation for any geometry and either linear or
non-linear behaviour would give%
\begin{equation}
-\frac{dU_{\max}}{dN}=-\frac{dU_{\max}}{da}\frac{da}{dN}%
\end{equation}
so if we attempt a power law fit
\begin{equation}
\frac{da}{dN}=C_{1}\left(  -\frac{dU_{\max}}{dN}\right)  ^{m_{1}}=C_{1}\left(
-\frac{dU_{\max}}{da}\frac{da}{dN}\right)  ^{m_{1}}%
\end{equation}
we are really writing
\begin{equation}
\frac{da}{dN}=C_{1}^{\frac{1}{1-m_{1}}}\left(  -\frac{dU_{\max}}{da}\right)
^{\frac{m_{1}}{1-m_{1}}}%
\end{equation}

But from the definition of SERR $G_{\max}=-\frac{dU_{\max}}{da}$. Hence, if
\begin{equation}
\frac{da}{dN}=C_{2}G_{\max}^{m_{2}}=C_{2}\left(  -\frac{dU_{\max}}{da}\right)
^{m_{2}}%
\end{equation}
it results the Delft proposal is perfectly equivalent to the classical SERR
one, and the correspondence between the two power law is precisely that
\begin{equation}
C_{2}=C_{1}^{\frac{1}{1-m_{1}}};m_{2}=\frac{m_{1}}{1-m_{1}}%
\end{equation}

Figure 1 shows the fatigue resistance curves plotted as $\frac{da}{dN}%
=C_{2}G_{\max}^{m_{2}}$ with different pre-crack lengths for Specimen F\_2
(45//45 interface) of the Yao et al(2014) paper, showing the data converge for
large precrack size with a $m_{2}=11.2$, but for the small precrack the
exponent is also rather different. Fig.2 shows the same data but in terms of
$\ \frac{da}{dN}=C_{1}\left(  -\frac{dU_{\max}}{dN}\right)  ^{m_{1}}$ show a
$m_{1}=0.8.$ Hence, with our prediction we should have $m_{2}=\frac{0.8}%
{0.2}=4$ --- the general trend we predict is correct, although we have made
the calculation very simple with assuming $R=0$ which is not really their case
of $R=0.5$.

\begin{center}%
\begin{tabular}
[c]{l}%
\centering\includegraphics[height=65mm]{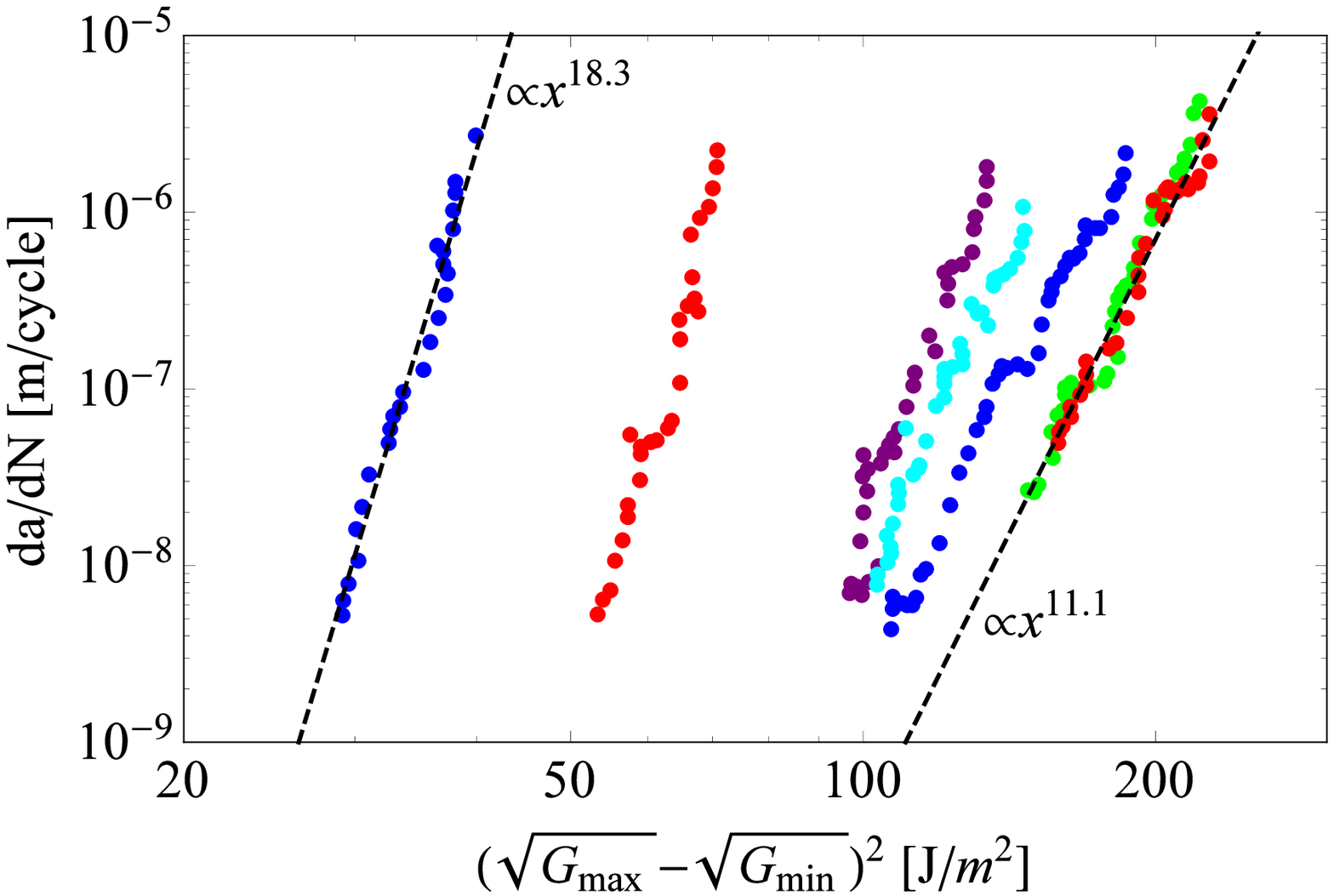}
\end{tabular}

Fig.1 - The data from Yao et al. show $\frac{da}{dN}=C_{2}G_{\max}^{m_{2}}$
(is $G_{\min}=0$?) with $m_{2}$ varying from $m_{1}=18$ to $m_{1}=$11 when
pre-crack is long enough%

\begin{tabular}
[c]{l}%
\centering\includegraphics[height=65mm]{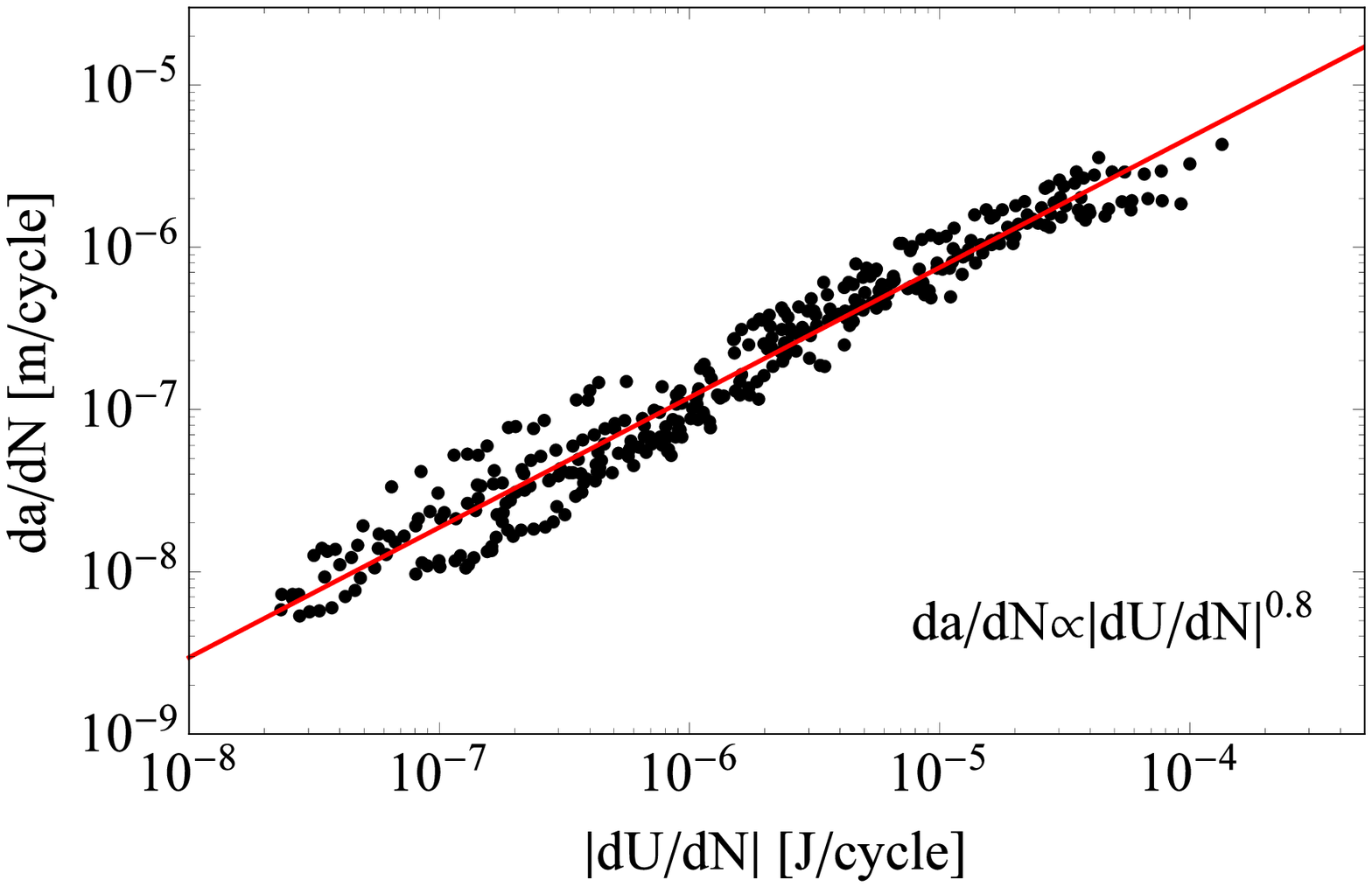}
\end{tabular}

Fig.2 - The data collapse for various laminates using $\frac{da}{dN}%
=C_{1}\left(  -\frac{dU_{\max}}{dN}\right)  ^{m_{1}}$ with $m_{1}=0.8$
\end{center}

\section{Discussion and non-linear models}

When dealing with crack fibre briding, we can not estimate $G$ using LEFM. As
Barenblatt and Botvina clarified very clearly (see Ciavarella et al., 2008),
clear dimensional analysis arguments elucidate that Paris' power-law is a weak
form of scaling, and Paris' parameters $C$ and $m$ should not be taken as true
"material constants". Indeed, they are expected to depend on all the
dimensionless parameters of the problem, and can be considered close to be
\textquotedblleft constants\textquotedblright\ only for example when crack
length is much larger than any other intrinsic characteristic of the problem,
including material length scales -- for a cohesive zone model, the length of
it should be much smaller than the crack size. Size-scale dependencies of $m$
and $C$ like those reported by Yao et al(2014) were known much earlier in
concrete in his studies of concrete (see see Ciavarella et al., 2008).

We attempt here a very preliminary and qualitative use of the
Barenblatt-Dugdale cohesive model for a crack from $x=0$ to $x=a$, for which
we have a "process zone" of size $b$ and this is such to eliminate the stress
singularity at the fictitious crack tip, $x=a+b$. \ Therefore, there are two
SIFs, one due to external loading, and one due to cohesive stresses
$\sigma_{c}$ (like fibre bridging stresses, or yield stress in Dugdale
material, or true cohesive stress in the Barenblatt original model). $G$ can
then be computed as
\[
G=\sigma_{c}\delta
\]
where $\delta$ is the COD at the mouth of the real crack $x=a$, i.e. at the
end of the cohesive zone. It is clear that for the SIF corresponding to the
external loading, the linear equations apply. But the cohesive stress region
introduces a non-linearity, the bigger the cohesive region is. And the true
$G$ is less than that computed from the external loading condition only. When
a crack develops a large cohesive zone compared to its size, ie. when the
crack is small, we have the bigger deviation from linearity, whereas when the
crack is sufficiently long, the cohesive zone becomes small enough for SSY to
occur and the cohesive model converges essentially to the LEFM linear predictions.

Delamination with fibre bridging shows the typical elastic-plastic fracture
mechanics Irwin's crack extension resistance curve (R-curve). This is why for
larger precracks, Yao et al. find much lower Paris curves coefficient $C$.
However, R-curve depends on the geometry and $G$ may be difficult to
calculate. Also, Yao et al (2014) discuss, the cohesive model would need to
predict a different cohesive zone size in static and fatigue condition for the
same crack size.

For illustrative purposes, let us consider the Barenblatt-Maugis model the
length of this cohesive zone for a crack in a infinite sheet (Cornetti et
al.,2016) is, under SSY\ (Small Scale Yielding) ($b<<a$)
\begin{equation}
b=\frac{\pi}{8}\left(  \frac{K_{I}}{\sigma_{c}}\right)  ^{2} \label{b}%
\end{equation}
while $G=\sigma_{c}\delta$. We retain however at least a second order term in
the COD (Crack Opening Displacement)
\begin{equation}
\delta=\frac{8\sigma_{c}a}{\pi E}\ln\left(  \frac{a+b}{a}\right)  \simeq
\frac{8\sigma_{c}a}{\pi E}\left(  \frac{b}{a}-\frac{b^{2}}{2a^{2}}\right)
\end{equation}

Then,
\begin{equation}
G=\frac{K_{I}^{2}}{E}\left(  1-\frac{\pi}{16}\frac{K_{I}^{2}}{\sigma_{c}^{2}%
a}\right)  =G_{el}\left(  1-\frac{\pi}{16}\frac{EG_{el}}{\sigma_{c}^{2}%
a}\right)
\end{equation}
The first term is the classical linear term, while the second term gives a
reduction which depends on cohesive stress, and size of the crack ---
disappearing for large cracks (for which there is no "bridging" in our
composite delamination case). Therefore, a Paris law in this case could be
written as
\begin{equation}
\frac{da}{dN}=C_{3}G_{\max}^{m_{3}}=C_{3}\left(  1-\frac{\pi}{16}%
\frac{EG_{el,\max}}{\sigma_{c}^{2}a}\right)  ^{m_{3}}G_{el,\max}^{m_{3}%
}=C_{3,eff}G_{el,\max}^{m_{3}}%
\end{equation}
so $C_{3,eff}=C_{3}\left(  1-\frac{\pi}{16}\frac{EG_{el,\max}}{\sigma_{c}%
^{2}a}\right)  ^{m_{3}}$ would be reduced for shorter cracks, as it appears
indeed in the data of Yao et al.(2014) when they plot Paris plots in terms of
$G_{el,\max}^{m_{3}}$.

Cohesive models for fatigue delamination usually are of different type, they
try to model the damage occurrying during fatigue see (Pascoe et al., 2013),
and we are not aware of a simple model like the one we are describing. But it
may well predict more closely the results Yao et al.(2014) seem to observe.

\section{Conclusion}

The proposal by Yao et al.(2014) to compute the crack driving force in fatigue
from the actual work done in the loading testing machine is not in contrast
with a Strain Energy Release Rate (SERR) approach, but only different from the
SERR computed from a linear model. In the classical problem where we have no
deviations from a Small Scale Yielding (SSY) assumption, we have derived the
relationship between the Yao et al.(2014) Paris-type proposal and the
classical Paris type equation on $G_{\max}$. \ The observation from the group
of Benedictus is anyway extremely simple and useful. It clarifies important
aspects of fatigue crack propagation, with particular reference to materials
with R-curve behaviour like delamination in composites. There are some not
very clear points in the paper by Yao et al.(2014) who at some point seem to
confuse an energy balance approach like in Griffith fracture, with a fatigue
process which is only a weak form of scaling with the Irwin Stress Intensity
Factor (at least in the original Paris form), or in terms of more general
crack driving forces, in later generalizations. The main drawback of the Delft
proposal is perhaps that it is not predictive, as it requires the actual
measurement during the fatigue process of the work input in the specimen by
the testing machine. It rather suggest, instead, that cohesive models which
appear not to have been attempted in a simple form as we are discussing here,
would be more successful than LEFM models in predicting fatigue delamination
crack growth.

\section{References}

Ciavarella, M., Paggi, M., \& Carpinteri, A. (2008). One, no one, and one
hundred thousand crack propagation laws: a generalized Barenblatt and Botvina
dimensional analysis approach to fatigue crack growth. Journal of the
Mechanics and Physics of Solids, 56(12), 3416-3432.

Cornetti, P., Sapora, A., \& Carpinteri, A. (2016). Short cracks and
V-notches: finite fracture mechanics vs. cohesive crack model. Engineering
Fracture Mechanics, 168, 2-12.

Pascoe, J. A., Alderliesten, R. C., \& Benedictus, R. (2013). Methods for the
prediction of fatigue delamination growth in composites and adhesive bonds--a
critical review. Engineering Fracture Mechanics, 112, 72-96.

Yao, L., Alderliesten, R. C., Zhao, M., \& Benedictus, R. (2014). Discussion
on the use of the strain energy release rate for fatigue delamination
characterization. Composites Part A: Applied Science and Manufacturing, 66, 65-72.

\end{document}